\documentclass[submission,copyright,creativecommons,colorlinks]{eptcs}
\providecommand{\event}{Lococo 2011} 
\usepackage[utf8]{inputenc}
\usepackage{graphicx}
\usepackage{multirow}

\title{Sources of Inter-package Conflicts in Debian}
\author{Cyrille Artho \qquad\qquad Kuniyasu Suzaki
\institute{Research Center for Information Security\\AIST, Tsukuba, Japan}
\email{\quad c.artho@aist.go.jp \qquad k.suzaki@aist.go.jp}
\and
Roberto Di Cosmo \qquad\qquad Stefano Zacchiroli
\institute{Laboratoire PPS\\
Université Paris Diderot -- Paris 7, France}
\email{\quad roberto@dicosmo.org \quad\qquad zack@pps.jussieu.fr}
}
\def\titlerunning{Inter-package Conflicts in Debian}
\def\authorrunning{C. Artho et al.}
\begin{document}
\maketitle

\begin{abstract}
Inter-package conflicts require the presence of two or more packages
in a particular configuration, and thus tend to be harder to detect
and localize than conventional (intra-package) defects. Hundreds of
such inter-package conflicts go undetected by the normal testing and
distribution process until they are later reported by a user.
The reason for this is that current meta-data is not fine-grained
and accurate enough to cover all common types of conflicts.
A case study of inter-package conflicts in Debian has shown that with
more detailed package meta-data, at least one third of all package
conflicts could be prevented relatively easily, while another one third could be
found by targeted testing of packages that share common resources or
characteristics. This paper reports the case study and proposes ideas
to detect inter-package conflicts in the future.
\end{abstract}

\section{Introduction}

\subsection{Package-based software distributions}

Modern software distributions are organized into packages. A software
package is a self-contained unit that can be installed or removed
independently of other packages, as long as dependencies are met.
A package manager controls such administrative tasks; compared to
unmanaged installations, the benefits of a package-based approach
are the ability to automatically install, upgrade, and remove packages
without the need to remember installation locations or which files
are affected by a change.

In real software, this ideal state is not easy to achieve, due to
dependencies between software packages, and interactions between
software belonging to different packages. Dependencies arise because
some packages provide lower-level functionality used by others.
Interactions occur on shared resources, such as files, and because
packages may provide components that can be combined into a larger
system (such as client and server packages communicating together).

Dependencies restrict the ability to freely install, remove, or
upgrade packages. If a package $a$ depends on another package $b$,
a package manager automatically requires $b$ to be installed when
$a$ is requested to be installed. Furthermore, package $b$ cannot
be removed as long as $a$ is still in use. Finally, upgrades of
one package often require a simultaneous upgrade of related packages.
In addition to this, there is a notion of \emph{conflicting} packages:
two packages may use the same resource or provide the same service
in a way that is incompatible with each other, so only one of these
two packages may reside on a system at any given time.

In package-based software distributions, so-called
\emph{package meta-data} describes dependencies and relations
between packages. Most Free Open Source Software (FOSS) systems
are managed in that way. Meta-data contains information about
dependencies of packages, and conflicts between them. At the time
of writing, meta-data covers relations between packages on a package
level; dependencies and conflicts are indicated by package, not
by the actual resources a package provides or depends on. So-called
\emph{virtual packages} are sometimes used as place-holders for
actual resources or services provided by a package, but they do
not constitute an accurate, fine-grained description of those
resources, which may be files, network ports, or system services.

\subsection{Inter-package conflicts}

Inter-package conflicts occur if the combination of multiple
packages results in a defect that is absent otherwise.
Package meta-data may indicate such conflicts, which prevents
conflicting combinations of packages from being installed.
However, inter-package conflicts may still arise in practice.
The reasons for such
conflicts are manifold: Packages are not simply bundles of files,
but include pre-installation and post-installation scripts. These scripts
are unrestricted, Turing-complete programs running with full system
(root) access. It is therefore impossible in general to capture
the full side effects of these scripts with a formal description.
The same problem arises of course as well for executing the software
provided by these packages. Therefore, a complete logical analysis
of package behavior is not possible; however, as this paper shows,
steps can be taken towards covering certain types of common
conflicts that are not automatically verifiable with current tools.

Another problem arises from the fact that meta-data is provided
manually, by package maintainers. It is therefore a challenge to
keep such meta-data up to date and accurate. This challenge
becomes especially daunting in the presence of a huge number of
software packages in distributions such as Debian GNU/Linux,
where the number of packages available currently exceeds
30,000~\cite{debian-sid-packages}.

As a consequence of this, bug reports referring to conflicts
between packages are becoming frequent. This paper investigates
the sources of the conflicts and tries to answer the following
questions:

\begin{itemize}
\item What are the reasons why inter-package conflicts arise?

\item Are there common categories of inter-package conflicts?

\item Can these problems be addressed by using existing tools,
or is there a need to augment existing tools, or create new ones?

\item Is the package meta-data currently being used, accurate
and sufficient? Is there a need to automatically verify such
meta-data for accuracy, or is there a need to use additional
meta-data for a more accurate notion of package conflicts? In
other words, are most or all possible package conflicts covered
by meta-data?
\end{itemize}

This paper is organized as follows: Section~\ref{sec:related-work}
describes related work. Section~\ref{sec:evaluation} shows a case
study on inter-package conflicts in Debian, with a detailed
evaluation of different kinds of package conflicts.
Section~\ref{sec:discussion} discusses the results and proposes
possible strategies for remedying problems found, and
Section~\ref{sec:conclusions} concludes and outlines future work.

\section{Related Work\label{sec:related-work}}

\subsection{Software packaging}

Software packages are a well-known example of the component models that have
originated from the field of component-based software engineering
(CBSE)~\cite{clemensbook, brown:cbse-state}. Packages fit very well within
common component definitions, but the raise in their popularity---started with
the advent of FOSS package managers such as the FreeBSD porting
system~\cite{stokely:freebsd}, APT~\cite{apt-howto}, Yum, etc.---has
highlighted very specific challenges related to their
deployment~\cite{hotswup:package-upgrades}. Some of those challenges are being
addressed relying on package meta-data and their formalization.

Seminal work by Mancinelli et.~al~\cite{edos2006ase} has shown how to encode
the installability problem for software packages as a SAT problem, established
the (NP-Hard) complexity of the problem, and shown applications of the encoding
to improve the quality of package repositories by avoiding non installable
packages. Based on the same formalization, various quality metrics have been
established, such as strong dependency and
sensitivity~\cite{strongdeps-esem-2009} (to evaluate the ``importance'' of a
package in a given repository) and strong
conflicts~\cite{dicosmo:strongconflicts} (to pinpoint packages which might
hinder the installation of several other packages). In the same vein, package
meta-data have also been used to predict future (non) installability of software
packages~\cite{abate:prediction}. The abundance of studies that rely on
package meta-data testifies the importance of the correctness of those
meta-data.

On the other hand, studies on package meta-data correctness like this one, seem
to be scarce. At the same time, a few testing can be found in the realm of
Quality Assurance (QA) of FOSS distributions to discover \emph{symptoms} that
might then lead, a human, to discover errors in package meta-data. To name one,
the ``file overwrite''~\cite{debian-file-overwrites}
initiative by Treinen helps in discovering undeclared conflicts among packages
in the Debian distribution.

\subsection{Alternatives to globally managed software packaging}

As an alternative to globally managed software packages that are organized in a
fine-grained hierarchy, self-contained packages including all sub-components,
sometimes called \emph{bundles}, are sometimes used. Such bundles include the
application and all libraries it depends on, linked
statically~\cite{presser-linkers}.  This contrasts to FOSS distributions where
libraries are shared, and generally required to be shipped as separate
packages---see for instance~\cite{debian-policy}, ``convenience copies of
code''---in order to ease the deployment of (security) upgrades.
In a system using bundled software, all applications using the library
in question need to be updated separately. This usually entails a
longer period during which a system is vulnerable, because some
software bundles may be provided by third parties.

An advantage of self-contained software bundles is the ease of testing
and deployment, as system-specific configurations and libraries have
only limited impact on the software bundle. However, statically linking
all libraries used by a bundle requires much disk space.
If many applications include the same statically-linked libraries,
these libraries are duplicated within the same system. Deduplication
addresses this problem~\cite{Collberg05slinky,Suzaki_movingfrom}.
Memory and storage deduplication merge same-contents chunks
on block level, and reduce the consumptions of physical memory.
By sharing identical chunks of storage, logical-level redundancies
caused by static linking are resolved on the physical level.

\section{Evaluation of Inter-package Conflicts\label{sec:evaluation}}

\subsection{Methodology}

The evaluation of existing inter-package conflicts in Debian
was carried out on a snapshot of the Ultimate Debian Database
(UDD)~\cite{udd-msr10}. This database contains key data of all active (open)
bugs at that time, such as bug ID, title, and the package involved.
The snapshot used was taken on January 23 2011, and contained
79936 bugs.

This database is too large to be analyzed manually, so the selection
of bugs was first narrowed down by a keyword search. We chose three
keywords to search for: ``break'', ``conflict'', ``overwrite''. The
first two words are generic descriptions of inter-package conflicts
and often appear in the form ``$a$ breaks $b$'' or ``$a$ conflicts
with ``$b$''. The last keyword describes one of the most common
inter-package problems, where one package overwrite a resource needed
by another package.

Table~\ref{table:keyword-matches} gives an overview of all the matches
in the search. A total of 929 bugs match the initial search; some
of the matches contain more than one keyword and therefore are
duplicates. Our aim is not to get an exact number of how many
inter-package conflicts there are in total; rather, we want to know what
types of conflicts occur more often than others, relative to the total
number.

\begin{table}
\begin{center}
\begin{tabular}{lrr}

Keyword&Matches&Refined matches\\\hline

break&575&161\\
conflict&252&85\\
overwrite&102&44
\end{tabular}
\end{center}
\caption{Number of matches per keyword in Debian database.\label{table:keyword-matches}}
\end{table}

We therefore narrow down the search to eliminate bug reports
that describe problems that relate to one package alone, rather than
a conflict between two packages. For example, ``overwrite'' could
appear in a bug report on a text editor in a bug report related to
 overwriting text (or a file, on saving). In fact, an initial manual
evaluation showed that about half of all bug reports found in the
initial search were not related to inter-package conflicts.
The search is refined to include only bug reports out of the initial
selection, where the title contains the name of another package.
This may filter out more bug reports than necessary (decreasing recall,
in search terms), but makes the results much more precise. To
avoid excluding too many packages, (version) numbers of packages are
not included in this filter, even if the package name itself contains
a version number. A manual check showed that this filter was actually a
good approximation of a manual selection of true inter-package conflicts.

As shown in Table~\ref{table:keyword-matches}, the refined selection
contains 290 matches, 241 of which are distinct bug reports. Further
manual post-processing of that list removes another 51 items, where
the title indicates clearly that those are not inter-package conflicts.
This leaves 190 bug reports where, judging from the title of the report,
a possible inter-package conflict is reported.

A subset of these bug reports was evaluated in a first sample, to come
up with a categorization of bug reports that would not be too coarse
(giving only a few rough classes of bugs) and not be overly precise
either (putting most bugs into a category of their own). After that,
all bug reports are classified according to these criteria, or
eliminated as not being inter-package conflicts, although the title
would suggest so (in the list of 190 reports).

\subsection{Results}

The 190 cases of which the bug report titles suggested an inter-package
conflict, were analyzed manually. This requires the full information
available on each bug, which is not contained in the summary database
(UDD) used in the first step. The 190 bug reports in question were
downloaded from the web page at \url{http://www.debian.org/Bugs/}.
51 bug reports out of 190 
contain no inter-package conflict, but rather a conflict that is
not reproducible, or a conflict within a single package which is either
misclassified or contains a misleading title. This leaves 139 genuine
inter-package conflicts, which are classified into five broad categories:

\begin{enumerate}
\item Conflicts on files and similar shared resources (such as devices
or C library function names). Whenever a conflict occurs directly on a
file (or device), the conflict is caught at installation time by
\texttt{apt-get}, the package manager for Debian. This handling
is safe, but unsatisfactory: if a list of files used was provided
beforehand, then an enhanced package manager could prevent an
installation attempt that is bound to fail. On the other hand, other
types of conflicts, such as name clashes in libraries, may not be
detected until an application is used at run-time.

\item Conflicts on shared data, configuration information, or the
information flow between programs. Configuration information is often
found in \texttt{/etc}, while shared data may be located elsewhere.
Information flow refers to function calls or communication via
pipes or a network. There are two basic cases where conflicts occur
on data or communication:
(1) An installation action of a package changes the configuration
such that either the syntax of a configuration file is broken (made
unreadable for the parser used by another tool), or the semantics
create a conflict. (2) A change in the data format between versions
of an application, which requires updating other components; the
lack of an appropriate newer version of other components, or the lack
of a declaration of such, causes a conflict. In both cases (1) and (2),
the conflict only becomes evident at run-time.

\item Uncommon, previously untested combinations of packages, cause
a conflict. In some cases, a package $a$ using another package $b$
makes a previously undetected fault in $b$ evident; it is possible
that other use cases for $b$ could produce the same problem, so the
failure can (at least in theory) be reproduced using $b$ alone.
In other cases, the combination of $a$ and $b$ is necessary for those
packages to fail, and either package would work fine without the
conflicting package being present. Nine cases fit this description,
where the reason of a conflict could not be attributed to cases listed
above.

\item Package evolution issues. When a software distribution evolves,
packages may be renamed or split up into multiple packages, or several
packages may be merged into one. This may require updating meta-data
in other packages for the distribution to remain consistent.
Furthermore, version changes with a package may also require meta-data
changes due to possible incompatibilities mentioned above.
Unfortunately, meta-data changes are not automated, and are primarily
the responsibility of the maintainer of a given package. This causes
a potential for meta-data to be outdated and not reflect a correct
state anymore.

\item The last category represents cases where two packages are
incorrectly classified as conflicting, although there is no conflict,
at least not for the current version of these packages.
\end{enumerate}

\begin{table}
\begin{center}
\small{
\begin{tabular}{rrl}
\multicolumn{2}{l}{\# of conflicts}&Conflict type\\\hline

48&&access to files and similar shared resources\\
&22&package provides same file as other package\\
& 4&package installers modify or delete file used by other package\\
& 2&file missing that is supposed to be provided by other package\\
& 3&packages modify/disable same shared resource or package\\
& 3&file permission conflict on shared file\\
& 4&file/directory name conflict (for names including version number etc.)\\
& 9&clashing C library symbols/function names/device names\\
& 1&package removal script corrupts system\\\hline

47&&file/API/data/configuration format\\
&17&update/installation breaks configuration or file format\\
&14&package breaks on uncommon or user-defined configuration/setting\\
& 4&package use (post-install) overwrites/breaks configuration files\\
& 9&API change between different package version breaks other package\\
& 3&kernel package not compatible with given version of other package\\\hline

22&&rare (previously untested) combination of packages\\
&13&defect in one package made visible by installation/use of other package\\
& 9&uncommon combination of packages makes one or more packages always fails\\\hline

14&&package evolution (split/merge/change) or faulty meta-data results in conflict\\
& 9&incorrect/outdated dependency meta-data (requires/conflicts)\\
& 5&package renaming/split/merge results in incorrect meta-data of other package\\\hline

 8&&spurious ``conflicts'' declaration prevents compatible packages from being used\\\hline
\end{tabular}
}
\end{center}
\caption{Overview of all package conflicts found in the Debian bug database.\label{table:deb-pkg-conflicts}}
\end{table}

Table~\ref{table:deb-pkg-conflicts} and Figure~\ref{fig:xpkgbugs}
show an overview of the
classification into these five categories. Larger categories were
split up into smaller groups to get a more detailed picture.
While human error in the classification is possible, the results are
overall quite clear for larger categories. Some trends are evident:

\begin{figure}[p]
\begin{center}
\includegraphics[scale=.7]{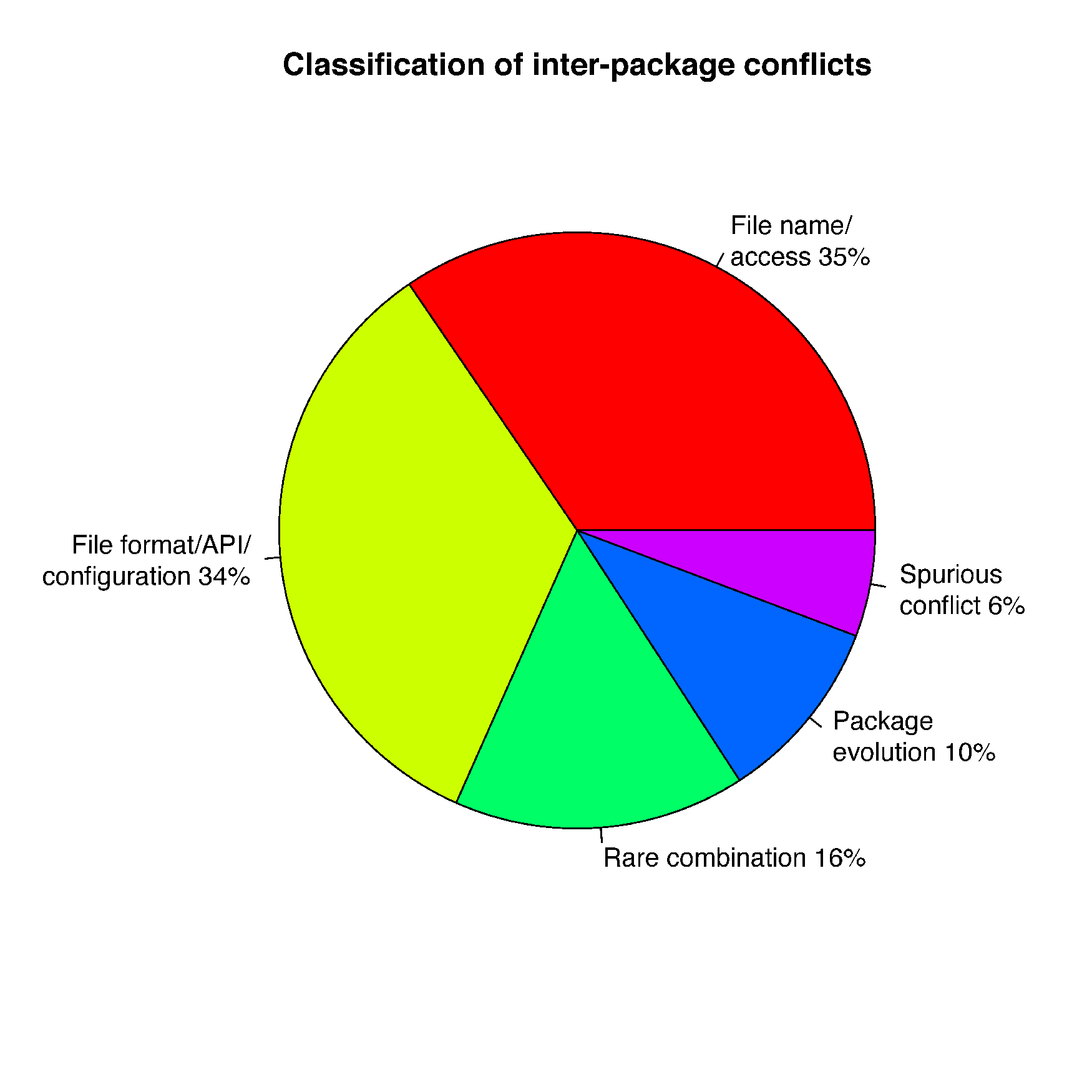}
\end{center}
\caption{Overview of sources of inter-package conflicts.\label{fig:xpkgbugs}}
\end{figure}



\begin{enumerate}
\item Resource conflicts are common, representing more than one third
of all conflicts. 22 out of 48 such conflicts are on files and caught
by the package manager at installation time; other similar conflicts may
not be caught until a package is actually used.

\item Conflicts on configuration, and to a lesser degree, the format of
shared data, are equally common. 17 cases were found where syntactic
problems caused a conflict between packages; the most common reason is the
automatic modification of configuration files by installation scripts.
These installation scripts are likely tested for common configurations,
but may not behave as expected for less common settings. Unintended
semantic changes in configuration files occurred 14 times during
installation, and 4 times after installation, so this is also a
significant problem. It is compounded by the fact that many files
have to be customized by the user before a package can be used, and
the formatting of a configuration file may see subtle changes that
are correctly dealt with by the packaged software itself, but not by
the installation scripts that manage the package.

\item Other problems between packages that are usually not installed
together represent one out of six inter-package conflicts. The huge
number of available packages makes it impossible to test all
combinations (or even just all pairwise possible combinations) of
packages together, so a conflict often goes undetected until reported
by a user.

\item Conflicts on meta-data level, often caused by package evolution,
contribute about 10\,\%.

\item Incorrect (or outdated) information on conflicting packages
sometimes occurs as well, which does not create a package conflict per
se, but instead prevents two packages from being used together.
\end{enumerate}

\section{Discussion\label{sec:discussion}}

The previous section has given a categorization of inter-package
conflicts based on empirical data. We now propose possible solutions
that can potentially cover some or all instances of each class of
conflicts.

\begin{enumerate}
\item Conflicts on files are not directly covered by existing meta-data,
although they may be implied by package-level conflicts. Work is in
progress to systematically test package installations against overwriting
files provided by another package~\cite{debian-file-overwrites}. As an
alternative to this, \emph{file diversions} enable a package to install
files at a different location; work is in progress to automate
this.%
\footnote{\url{http://wiki.debian.org/SummerOfCode2011/DeclarativeDiversions},
  retrieved June 2011}

This case study shows that while the majority of such conflicts occurs
at file level, file \emph{permissions} (and ownership) rather than just
file names, and possible file/directory renaming actions during package
upgrades, should also be considered. Finally, coverage of similar resources such
as network ports and C function names would further augment the ability
of such tools to detect conflicts proactively.

More detailed meta-data will require much more space than existing
(rather compact) package meta-data. We propose that such extra meta-data
is generated and used only by developers and package maintainers.
As it covers possible conflicts proactively, at development time,
extra meta-data does not have to be included in the final distribution.
We think that most or all of such resource-related meta-data can be
extracted automatically from test runs, therefore requiring no extra
effort from package maintainers.

\item Conflicts on configuration files, file formats and API versions
are also common, and clearly demonstrate the need of systematic testing
against such conflicts. In the light of testing against overwriting
files~\cite{debian-file-overwrites}, inter-package tests should also
be automatically run against conflicts on shared data. This is much
more difficult to automate, and only feasible for packages that
include automatic regression tests.

The problem is that regression tests are used
by developers and package maintainers, but not by end users who
install and use these packages. Because of this, regression tests
are currently not covered by package meta-data. This makes them
inaccessible to today's package management tools, and pretty
much precludes the automated discovery of such intricate conflicts.
However, at a lower level, many source-level distributions have
a ``make test'' or ``make check'' build target that automatically
performs such tests. In the future, such information could be provided
in package meta-data, for package maintainers. Furthermore, on a
basic level, certain problems may be found just by executing a program
and checking whether its return value indicates an error, or by
attempting to start and stop a system service cleanly.

\item The fact that rare combinations of packages may cause problems
is not surprising, given the large number of packages available.
An exhaustive testing of package combinations is not feasible, but
heuristic-based testing of sets of packages may be. A possible
approach may be to install larger subsets of packages, and to narrow
down the set of conflicting packages by a systematic search such
as delta debugging~\cite{zeller02simplifying}.

\item Package evolution often brings with it an invalidation of
package meta-data. The fact that about one tenth of inter-package
conflicts occurred directly due to invalid meta-data after larger
package modifications (such as splitting a package into two
packages), shows that meta-data needs to be verified for
consistency and accuracy. Especially when given a situation with
``known good'' meta-data (before the modification), automatic
verification of the new meta-data is feasible if packages can be
tested automatically.

As with other issues described above, meta-data does not cover
the requirements of packages in enough detail. For example, take
a package $a$ that is split up into $a'$ and $a''$, because some parts
of $a$ are not used by many packages. If a package $b$ depended on $a$
in the old configuration, it is possible that it will depend on $a'$,
$a''$, or both packages in the new configuration. If some of the
resources provided by these packages are loaded dynamically by $b$
(at run-time), then verification of the actual software is required
to determine the correct new dependency.

\item Spurious (or outdated) declarations of inter-package conflicts
can be responded to, by automated testing of packages that supposedly
conflict. As mentioned above, work is in progress to detect file-level
conflicts, but other types of conflicts require more detailed meta-data,
or mechanisms to better support the automatic testing of the execution
of the software that packages provide.

\end{enumerate}

\section{Conclusions and Future Work\label{sec:conclusions}}

Conflicts between software packages occur due to a variety of reasons.
Conflicts on shared resources and configuration files are particularly
common. The underlying problem is that package behavior at installation,
use, and de-installation time is unrestricted, so a complete formal
description of package behavior cannot be achieved. However, steps
can be taken towards increasing the expressiveness and accuracy of
package meta-data, by adding meta-data that is intended for package
developers and maintainers.

In our case study, we categorize a large number of
inter-package conflicts, and propose possible solutions to common
categories of conflicts.
Our study uses a single snapshot of bugs between packages reported in
Debian GNU/Linux. Future work includes studying the evolution of
packages, and bugs reported, in more depth by investigating multiple
snapshots over time. Furthermore, other software distributions such
as Fedora may also be considered.

As a conclusion from our initial case study, we found that ongoing and
future projects can reduce inter-package conflicts most efficiently by
(a) identifying and testing combinations of packages that may conflict,
(b) generating and using extra meta-data, and (c) checking the validity
of (manually provided) meta-data. Such meta-data should cover files
including file meta-data in particular, and as a next step, other
system resources such as network ports, shared (global) configuration
data, and communication between components. Another aspect currently
omitted in meta-data is information about regression tests that
already exist in many packages, but are inaccessible on a package
level because they are not declared or available in a uniform way.


\end{document}